\documentclass[preprint]{aastex63}

\usepackage{graphicx}

\begin{document}
%\title{A Search for a Simultaneous Optical Counterpart to FRB 20181130B with GWAC high cadence images}
\title{Constrains on optical emission of FAST-detected FRB 20181130B with GWAC synchronized observations}

\correspondingauthor{Liping Xin}
\email{xlp@nao.cas.cn}

\author{L. P. Xin}
\affiliation{CAS Key Laboratory of Space Astronomy and Technology, National Astronomical Observatories, Chinese Academy of Sciences, Beijing 100101, China.}

\author{H. L. Li}
\affiliation{CAS Key Laboratory of Space Astronomy and Technology, National Astronomical Observatories, Chinese Academy of Sciences, Beijing 100101, China.}

\author{J. Wang}
\affiliation{Guangxi Key Laboratory for Relativistic Astrophysics, School of Physical Science and Technology, Guangxi University, Nanning 530004, China}
\affiliation{CAS Key Laboratory of Space Astronomy and Technology, National Astronomical Observatories, Chinese Academy of Sciences, Beijing 100101, China.}

\author{X. H. Han}
\affiliation{CAS Key Laboratory of Space Astronomy and Technology, National Astronomical Observatories, Chinese Academy of Sciences, Beijing 100101, China.}

%\author{Y. Xu}
%\affiliation{CAS Key Laboratory of Space Astronomy and Technology, National Astronomical Observatories, Chinese Academy of Sciences, Beijing 100101, China.}

%\author{X. M. Meng}
%\affiliation{CAS Key Laboratory of Space Astronomy and Technology, National Astronomical Observatories, Chinese Academy of Sciences, Beijing 100101, China.}

\author{Y. L. Qiu}
\affiliation{CAS Key Laboratory of Space Astronomy and Technology, National Astronomical Observatories, Chinese Academy of Sciences, Beijing 100101, China.}

\author{H. B. Cai}
\affiliation{CAS Key Laboratory of Space Astronomy and Technology, National Astronomical Observatories, Chinese Academy of Sciences, Beijing 100101, China.}

\author{C. H. Niu}
\affiliation{National Astronomical Observatories, Chinese Academy of Sciences, Beijing 100101, China.}

\author{X. M. Lu}
\affiliation{CAS Key Laboratory of Space Astronomy and Technology, National Astronomical Observatories, Chinese Academy of Sciences, Beijing 100101, China.}

\author{ E. W. Liang}
\affiliation{Guangxi Key Laboratory for Relativistic Astrophysics, School of Physical Science and Technology, Guangxi University, Nanning 530004, China}

\author{Z. G. Dai}
\affiliation{Department of Astronomy, University of Science and Technology of China, Hefei 230026, China}
\affiliation{School of Astronomy and Space Science, Nanjing University, Nanjing 210023, China}

%=====================

\author{X. G. Wang}
\affiliation{Guangxi Key Laboratory for Relativistic Astrophysics, School of Physical Science and Technology, Guangxi University, Nanning 530004, China}

\author{X. Y. Wang}
\affiliation{School of Astronomy and Space Science, Nanjing University, Nanjing 210023, China}
\affiliation{Key Laboratory of Modern Astronomy and Astrophysics (Nanjing University), Ministry of Education, Nanjing 210093, China}

\author{ L. Huang}
\affiliation{CAS Key Laboratory of Space Astronomy and Technology, National Astronomical Observatories, Chinese Academy of Sciences, Beijing 100101, China.}

\author{C. Wu}
\affiliation{CAS Key Laboratory of Space Astronomy and Technology, National Astronomical Observatories, Chinese Academy of Sciences, Beijing 100101, China.}

\author{G. W. Li}
\affiliation{CAS Key Laboratory of Space Astronomy and Technology, National Astronomical Observatories, Chinese Academy of Sciences, Beijing 100101, China.}

\author{Q. C. Feng}
\affiliation{CAS Key Laboratory of Space Astronomy and Technology, National Astronomical Observatories, Chinese Academy of Sciences, Beijing 100101, China.}

 \author{J. S. Deng}
 \affiliation{CAS Key Laboratory of Space Astronomy and Technology, National Astronomical Observatories, Chinese Academy of Sciences, Beijing 100101, China.}
 
 \affiliation{School of Astronomy and Space Science, University of Chinese Academy of Sciences, Beijing, China}

 \author{S. S. Sun}
 \affiliation{Guangxi Key Laboratory for Relativistic Astrophysics, School of Physical Science and Technology, Guangxi University, Nanning 530004, China}
\affiliation{CAS Key Laboratory of Space Astronomy and Technology, National Astronomical Observatories, Chinese Academy of Sciences, Beijing 100101, China.}
 \affiliation{School of Astronomy and Space Science, University of Chinese Academy of Sciences, Beijing, China}

  \author{ Y. G. Yang}
  \affiliation{School of Physics and Electronic Information, Huaibei Normal University, Huaibei 235000, China. }

\author{ J. Y. Wei}
\affiliation{CAS Key Laboratory of Space Astronomy and Technology, National Astronomical Observatories, Chinese Academy of Sciences, Beijing 100101, China.}

%=====================

\begin{abstract}

Multi-wavelength simultaneous observations are essential to the constraints on the origin of fast radio bursts (FRBs). However, it is a significant observational challenge due to the nature of FRBs as transients with a radio millisecond duration, which occur randomly in the sky regardless of time and position. Here, we report  the search for short-time fast optical bursts in the GWAC archived data associated with FRB 20181130B, which were detected by the Five Hundred Meter Spherical Radio Telescope (FAST) and recently reported.
No new credible sources were detected in all single GWAC images  with an exposure time of 10 s, including image with coverage  of the expected arrival time in optical wavelength  by taking the high dispersion measurements into account. Our results provide a limiting magnitude of 15.43$\pm0.04$ mag in R band, corresponding to a flux density of 1.66 Jy or 8.35 mag in AB system by assuming that the duration of the optical band is similar to that of the radio band of about 10 ms. This limiting magnitude makes the spectral index of $\alpha<0.367$ from optical to radio wavelength.  The possible existence of longer duration optical emission was also investigated with an upper limits  of 0.33 Jy (10.10 mag), 1.74 mJy (15.80  mag) and 0.16 mJy (18.39 mag) for the duration of 50 ms, 10 s and 6060 s, respectively. 
This undetected scenario could be partially attributed to the shallow detection capability, as well as the high inferred distance of FRB 20181130B and the low fluence in radio wavelength. 
The future detectability of optical flashes associated with nearby and bright FRBs are also discussed in this paper.

\end{abstract}

%% Keywords should appear after the \end{abstract} command. 
%% See the online documentation for the full list of available subject
%% keywords and the rules for their use.

%% From the front matter, we move on to the body of the paper.
%% Sections are demarcated by \section and \subsection, respectively.
%% Observe the use of the LaTeX \label
%% command after the \subsection to give a symbolic KEY to the
%% subsection for cross-referencing in a \ref command.
%% You can use LaTeX's \ref and \label commands to keep track of
%% cross-references to sections, equations, tables, and figures.
%% That way, if you change the order of any elements, LaTeX will
%% automatically renumber them.
%%
%% We recommend that authors also use the natbib \citep
%% and \citet commands to identify citations.  The citations are
%% tied to the reference list via symbolic KEYs. The KEY corresponds
%% to the KEY in the \bibitem in the reference list below. 

\section{Introduction} \label{sec:intro}

Fast Radio Bursts (FRBs) are  bright, cosmological origin, and millisecond-duration  bursts in radio wavelengths
(Lorimer et al. 2007; Thornton et a., 2013; Bassa et al., 2017; Macquart et al., 2020). 
After the discovery of the first FRB (Lorimer et al., 2007), 
a number of dedicated facilities have been conducted to search FRBs, such as the Parkes telescope (e.g., Bhandari et al., 2018),
the updated Molonglo Observatory Synthesis Telescope (UTMOST; e.g., Farah et al., 2018);
the Australian Square Kilometer Array Pathfinder (ASKAP, e.g., Shannon et al., 2018),
the Canadian Hydrogen Intensity Mapping Experiment (CHIME; The CHIME/FRB Collaboration et al., 2018),
the Deep Synoptic Array (DSA; Ravi et al., 2019; Kocz et al., 2019),
the Green Bank Telescope (Masui et al., 2019), 
the Arecibo (Spitler et al., 2014; Patel et al., 2018) 
and the Five-hundred-meter Aperture Spherical radio Telescope (FAST, Nan et al., 2011, Li et al., 2019).
All these efforts result in an 
increasing rate of  new FRB detections.
%\footnote{ https://www.wis-tns.org/ Petroff \& Yaron (2020) }
%\footnote{http://frbcat.org (Petroff et al., 2016)}. 
Among them, more than 20 repeating FRBs have been reported.
Particularly,
the physical origin of the repeating FRB 20121102A was identified to be with a low-metallicity star-forming dwarf galaxy at a redshift 0.19273 (Tendulkar et al., 2017; Bassa et al., 2017). 
FRB 20190523A was found to be associated to a more massive but low specific star-formation rate (Ravi et al., 2019).
The identification of the counterpart of the brightest radio bursts from SGR 1935+2154
as a magnetar in our Galaxy by HXMT (Li et al., 2021) and INTEGRAL (Mereghetti et al., 2020) with short-duration
X-ray bursts suggests 
that at least a friction of FRBs are connected with  
new-born magnetized neutron star  (e.g., Weltman \& Walters 2020; Zhang 2021).
More  bursts with similar characteristics needed to be detected in the future to confirm this conclusion.
Recently, a new large  sample with 535 FRBs was presented by CHIME/FRB Collaboration (2021) which were detected by the CHIME survey, including 61 bursts from 18 previously reported repeating sources and 474 one-off bursts.
Though an increasing catalog of theories and models was developing to explain the physical nature of FRBs (e.g.,see the review of Platts et al., 2019; Xiao et al., 2021), the origin of FRBs remains a mystery.

Multi-wavelength observations are expected to 
place constraints on the emission mechanism of FRBs (e.g., Nicastro et al., 2021).
A bright optical counterpart associated to an FRB is model dependent (e.g., Zhang 2017; Ghisellini \& Locatelli 2018; 
Beloborodov 2020;
Nicastro et al., 2021). 
However, 
considering that the estimated FRB event rate is as high as 3600 FRBs/sky/day 
at $>$0.63 Jy at 350 MHz and 5 ms burst width (Chawla et al. 2017),
one would expected to detect some astrophysical optical transients with millisecond timescale
 accompanying FRBs under some extremely conditions (e.g., Lyutikov \& Lorimer 2016, 2020; Yang et al., 2019; Platts et al., 2019; Margalit et al., 2020; Beloborodov 2020).
For example,   the single-zone inverse Compton (IC) scattering in the pulsar magnetosphere model (Yang et al., 2019) predicts a bright millisecond optical emission under extremely condition. Beloborodov (2020) predicted that a radio burst could turn into a bright optical flash with a timescale of $\leq$1 second when a repeating magnetic flare  from a young magnetar strikes the wind bubble in the tail of a previous flare.
 Gamma-ray bursts and their multiwavelength afterglow could also be detected accompanied by FRBs due to  double neutron star mergers (e.g., Totani 2013; Wang et al. 2016, 2018; Zhang 2014). 
Besides, there are also  models to predict long duration optical emission associated with FRBs, for example, 
 type Ia supernovae  were predicted to be related to FRBs with model of binary white dwarf merger  (Kashiyama et al., 2013). 
 A long duration optical emission associated FRBs may be produced by the two-zone IC scattering process (Yang et al., 2019).

On the other hand, 
considering the complex observational characteristics of FRBs occurring with a millisecond duration at
 random times and positions on the sky, it is challenge
 searching the optical emission simultaneously at the time when the FRBs is detected. 
No positive result have been reported in the literature  up to date (e.g., Richmond et al., 2019; Karpov et al., 2017; Hardy et al., 2017; 
Tingay \& Yang et al., 2019; Kilpatrick et al., 2021).
 As discussed by Chen et al.(2020), the detection probabilities could be increased by improving the detection ability or shortening the cadence or increasing the field of view or lengthening the total observation duration. 
In practical, there still are  some possibilities by chance that the same patch of sky could be covered by a very wide field-of-view optical telescope with a high temporal resolution and a radio telescope, when an FRB is detecting.
 
 After the first FAST discovery of fast radio burst, FRB 20181123B (Zhu et al., 2020),  recently three new FRBs detected by FAST were reported (Niu et al., 2021) during Commensal Radio Astronomy FAST survey (CRAFTS). 
 Among them,  FRB 20181130B was detected by FAST in M11 beam ID on 13:01:27.034 UT at 30th Nov. 2018 at high galactic latitude (Niu et al., 2021). 
 The duration is 9.52$^{+5.94}_{-5.08}$ ms. The observed peak flux density and the measured fluence are $\sim$20.6 mJy
 and 0.168 Jy ms, respectively. 
 The Dispersion Measure is 1705.5$\pm$6.5 pc cm$^{-3}$, corresponding to an estimated maximum redshift of z$\sim$1.83 and a  luminosity of $1.6\times10^{42}$ erg s$^{-1}$ (Niu et al., 2021)\footnote{However, one shall be noted that the real distance of the source is uncertain with a value being less than the estimated.}
 The best coordination is RA=00:39:07.85, DEC=19:24:31.7, J2000 (Niu et al., 2021).
 The uncertainty of this location was not presented in the literature (Niu et al., 2021).
 However,
  since one beam covers the sky area with a diameter of 2.5 arcmin (Jiang et al., 2020), which could be taken as a maximum value for the error circle for the search of its optical counterpart. 
The pointing error could be neglected since 
the  pointing errors of the 19-beam receiver in different sky
positions are less than 16 arcseconds  and the standard deviation of
pointing errors is 7.9 arcseconds (Jiang et al., 2020).
Thus, with the above consideration,
the value of 1.3 arcmin as the radius of the uncertainty of the localization is adopted in this work.

 During the burst of FRB 20181130B, it chanced that the same field was
 also being monitored by the Ground based Wide Angle Cameras (GWAC). The total observations for this field lasted for more than four hours with a cadence of 15 seconds. 
 This is a great opportunity to search for any short-duration optical emission in the images obtained by GWAC synchronized  observations.
 
In this paper, we report the search for the short duration optical emission in the GWAC data covering the FRB 20181130B burst time. The GWAC observation and data reduction related to FRB 20181130B is presented in Section 2. 
The results and the discussion are given in Section 3.  Summary will be presented in Section 4.

\section{GWAC Observations and data processing}

GWAC (Ground-based Wide Angle Cameras) system, as one of the main ground-based facilities of $\text{SVOM}$\footnote{SVOM is a China-France satellite mission dedicated to the detection and study of Gamma-ray bursts (GRBs)}
mission (Wei et al. 2016), is an optical transient survey located at Xinglong observatory, China.  
This system is aiming to detect various of short-duration astronomical events including the electromagnetic counterparts of gamma-ray bursts (Wei et al. 2016) and gravitational waves (Turpin et al. 2020) and stellar flares (Xin et al., 2021) by imaging the sky at a cadence of 15 seconds down to $R\sim$16.0 mag. 
A real-time pipeline for short duration transient alert system was developed in GWAC system, named as GWAC transient alert system. With this system, the method in real-time pipeline for GWAC data to search any short duration transients was  catalog crossmatching. All  candidates passing the filters would be followed by two 60cm optical telescopes (GWAC-F60A/B) within two minutes after the alerts, consequently confirmed or rejected automatically by another realtime pipeline developed for GWAC-F60A/B data. If one event was confirmed, an alert will be promptly send to the GWAC duty scientist and displayed in the GWAC transient webpage for more detailed validations (Xu et al., 2020).

For GWAC, each mount carries four JFoV cameras (call one unit  in the GWAC system).
For each GWAC JFoV, 
the effective aperture size and the f-ratio are 18cm and $f/1.2$,  respectively.
Each JFoV camera is equipped with a 4096$\times$4096 E2V back-illuminated CCD chip, giving a field of view of 150 deg$^2$ and a pixel scale of 11.7 arc seconds.
The total FoV for each unit is $\sim$ 600 deg$^2$.  
The wavelength range is from 0.5 to 0.85 $\mu m$. 
Currently, four GWAC units have been set up.  
During the survey, each unit was assigned to imaging a given grid continuously which is predefined for the whole sky.
Areas of sky at a Galactic latitude of $b<20^\circ$ as well as the grids near the Moon are set with  lower priorities  (Han et al., 2021), since the detection efficiency of any transient observation in these areas will be reduced by the higher star density or higher background noise, due to the low spatial resolution of GWAC.
The exposure time is 10 seconds and the readout time is 4.479 sec for each frame, making a survey cadence of about 15 sec.
The observation time system was synchronized in real-time using GPS  with an accuracy of 10 milliseconds.
More detailed information can be found in references (Wang et al., 2020; Xin et al., 2021; Han et al., 2021).

FRB 20181130B was detected (Niu et al., 2021) in the radio region at 13:01:27.034 UT (denoted as $T_{0}$) on Nov 30, 2018.  
During the burst, GWAC was operating in survey mode. A large area of sky  with about 2200 square degrees were monitored by four GWAC units. Thanks for the large field of view, the location of FRB 20181130B was coincidently covered by one sky grid which was monitored synchronously by the camera G043 in the \#4  unit. 
The total observation time lasted from 10:18:33.1 UT to 14:42:14.9 UT, covering the FRB outburst time. 
During the observations, there is no any prompt alert produced by GWAC transient alert system for any new optical source around this position.

\begin{figure}[]
 \centering
 \includegraphics[width=0.8\textwidth]{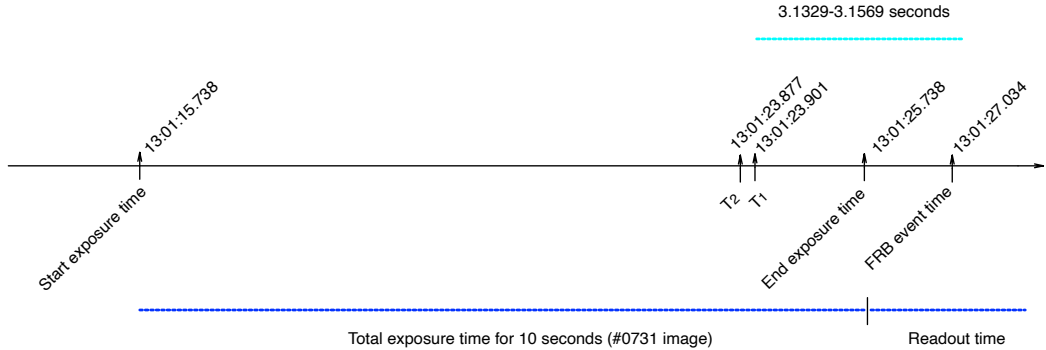}
 \caption{Cartoon figure to illustrate the time series of the start exposure time and the end exposure time of \#0731 image, the FRB event time and the time window between T$_1$ and T$_2$ for optical emission  derived by considering the uncertainty of the dispersion measurement. The blue dot line in the bottom shows the total exposure time and readout time. The green dot line in the top right presents the difference  between the arrive times in optical and radio wavelengths. All time is in UTC. 
  }
 \label{timeseries}
 \end{figure}

When FRB 20181130B was detected, the exposure time for  image  \#0731 obtained by G043 had just finished and the data was being read out from the camera. However, 
due to the large dispersion measurement of 1705.5$\pm$6.5 pc cm$^{-3}$ (Niu et al., 2021), 
the expected delay time between 
%the FAST radio burst  and the optical millisecond-scale emission could be estimated to be within 3.1-5.3 seconds\footnote{The detection frequency for the FAST is from 1.16 GHz to 1.5 GHz. When we estimate the time delay with the frequency of 1.5 GHz, the time delay between optical and radio would be 3.145 sec, while if 1.16 GHz is used for the estimation, the time delay would be 5.258 sec.}. 
the FAST radio burst  and the optical millisecond-scale emission was estimated to be within 3.1329-3.1569 seconds\footnote{The estimation is based on the reference of the arrive time at $\nu=1.5$ GHz} after considering the uncertainty of the dispersion measure (Niu et al., 2021).
In other words, 
it is possible that the associated optical emission was detected in  images taken before the radio burst  for this epoch, corresponding to a time window of 13:01:23.877  ($T_{2}$) and 13:01:23.901 UT ($T_{1}$), which is exactly covered by the effective exposure time of GWAC image \#0731 ( from 13:01:15:738 to 13:01:25:738 UT ). 
All these time series are displayed in Figure.\ref{timeseries}.

\begin{figure}[]
 \centering
 \includegraphics[width=0.8\textwidth]{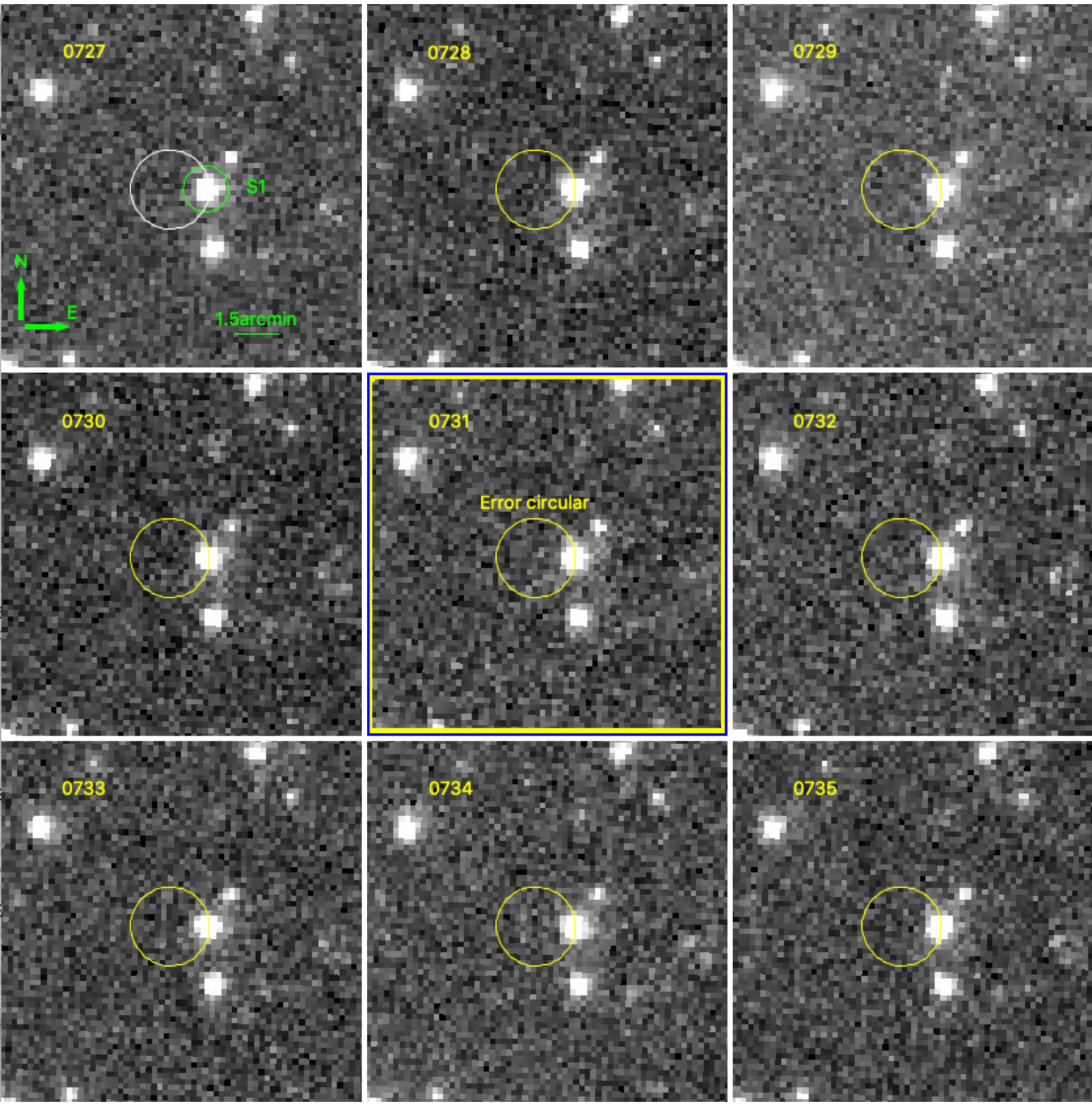}
 \caption{Finding Charts of FRB 20181130B detected by GWAC. All these images were acquired by GWAC on the same night,
 and all consecutive observation sequences are aligned and labeled. Each image is about 11.6$\times$11.6 arcmin in size with the north at the top and the east at the right. The image labeled with \#0731 was observed from 13:01:15.74 UT to 13:01:25.74 UT, completely covering the expected time of prompt optical emission $T_{opt}$. In image \#0731, there is a yellow circle showing the sky region where the position of FRB 20181130B is uncertain. The radius was set to 1.3 arcmin. The bright source labeled S1 in the image \#0727 is a Galactic  source identified from the Gaia DR2 catalogue with a G-band magnitude of 11.902 and a distance of $\sim$531 pc, and its association with FRB 20181130B can be ruled out. 
  }
 \label{findchart}
 \end{figure}

As shown in Figure.\ref{findchart},
nine consecutive images of GWAC around the burst time are displayed.
All these images are labeled with observation series number and aligned by each other.
 The size for each image is about 11.6$\times$11.6 arcmin with the north at the top and the east at the right.
A yellow circle with a radius of 1.3 arcmin in the center of each frame 
shows the sky region where the position of FRB 20181130B is uncertain. 
A very bright source nearby the error circle labeled as S1 is marked in the image \#0727, which is a cataloged  dwarf source  (J003902.54+192431.5) identified from the Gaia Dr2 catalog (Gaia Collaboration et al., 2018)
whose position is RA(J2000)=00:39:02.523, DEC(J2000)=+19:24:31.493. The G magnitude and the distance for S1  is 11.9022$\pm$0.0003 and 531.23142 pc, respectively. 
This object could be excluded for the association with  FRB 20181130B since the nature of the source.

\begin{figure}[htbp]
\begin{center}
 \includegraphics[width=0.5\textwidth]{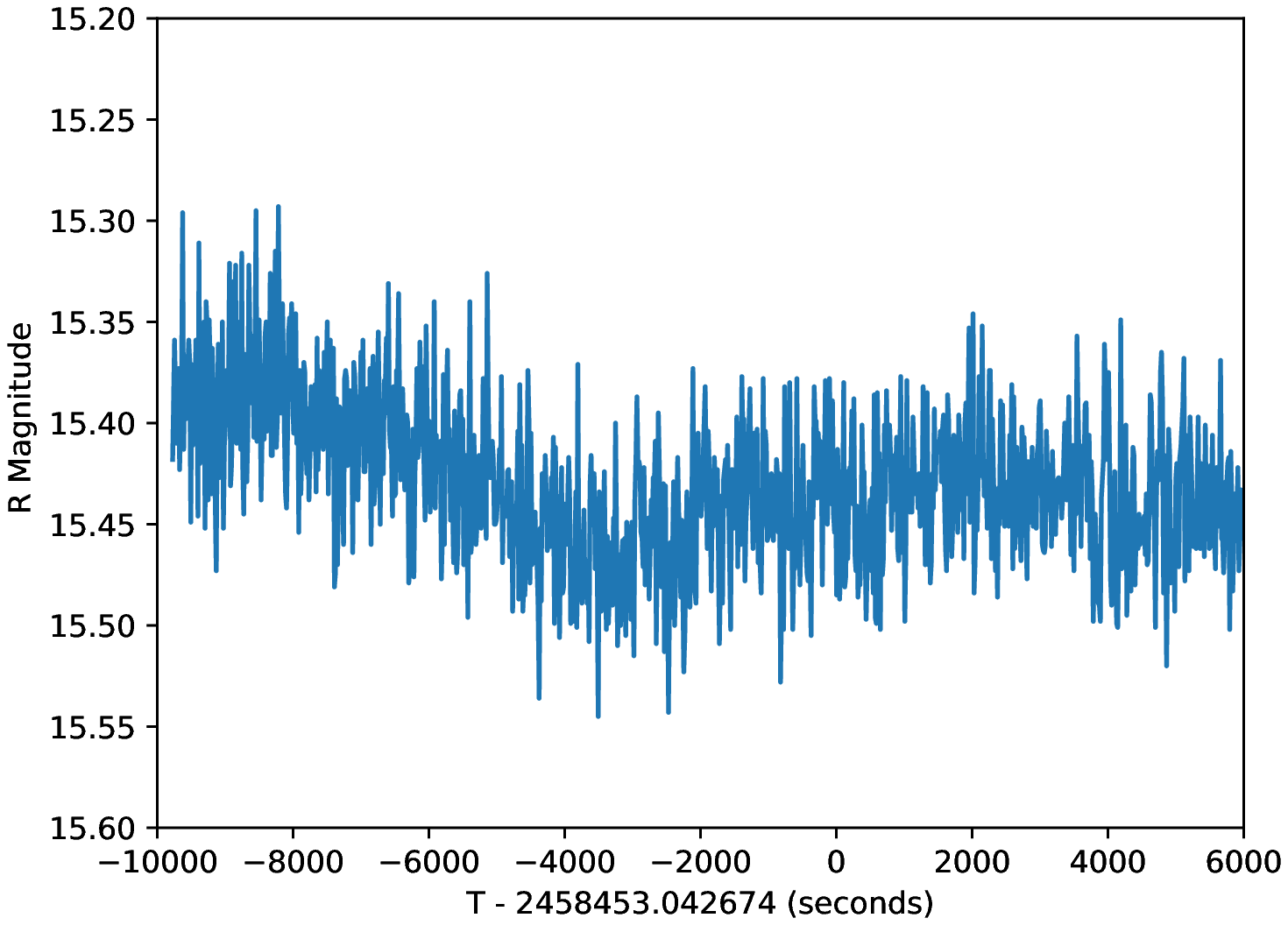}
  \includegraphics[width=0.5\textwidth]{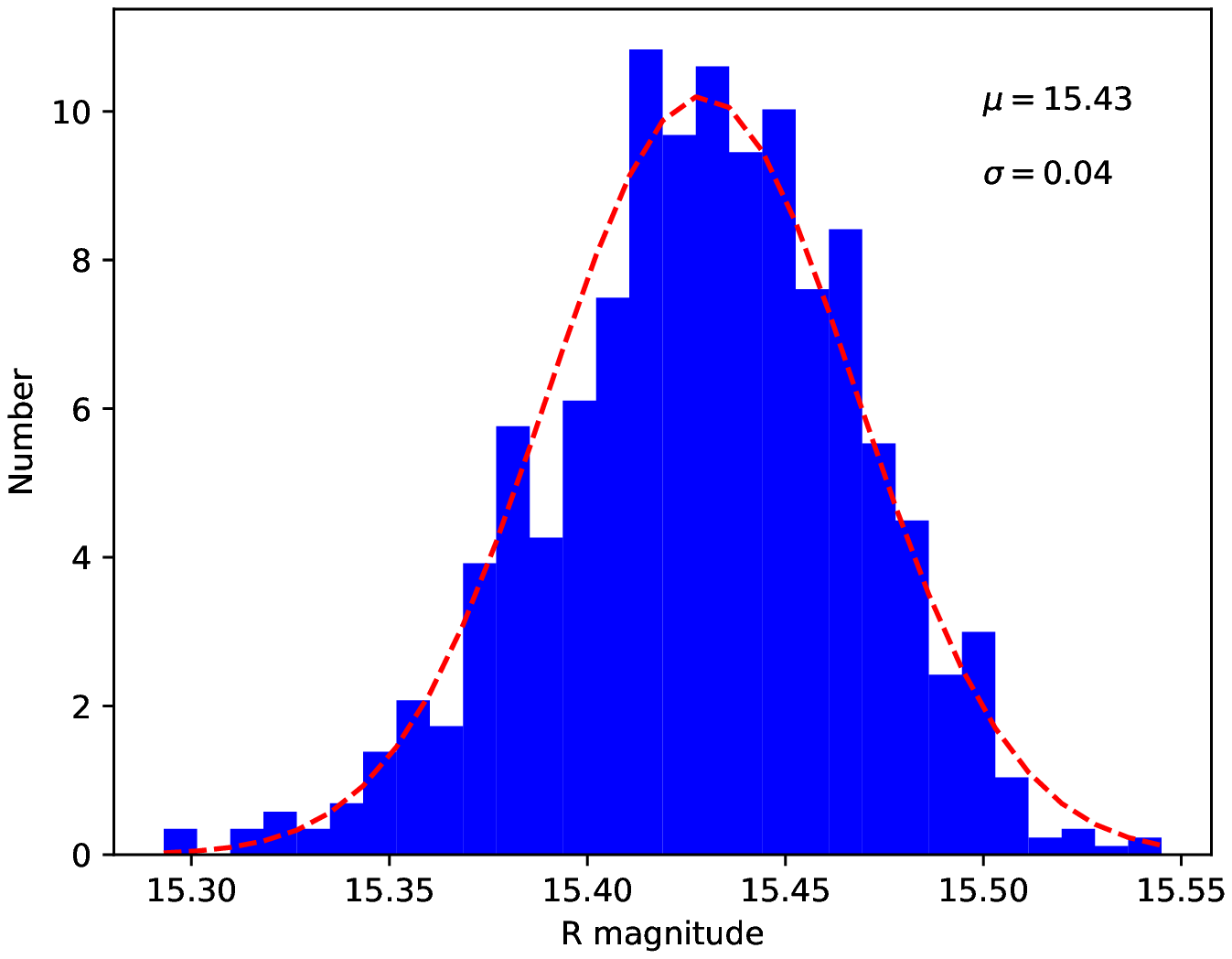}
\caption{Upper: The 3$\sigma$ upper limit magnitude of optical emission for FRB 20181130B derived from GWAC  data relative to the burst time (Niu et al., 2021) in seconds. The y-axis is the upper limit magnitude which was calibrated to USNO B1.0 catalog. Bellow: A statistics with a histogram plot in blue for the limit magnitude. The red dot line showed the fitting result with the gaussian distribution. The K-S test gave a p-value of 0.3303.}
\label{limitmag_lc}
\end{center}
\end{figure}

We performed an off-line pipeline to search the GWAC archived images for possible short-lived optical counterparts in around the time of this burst. 
First, all GWAC images have been corrected of bias, dark and flat-field in a standard manner using the IRAF\footnote{IRAF is distributed by the National Optical Astronomical Observatories, which are operated by the Association of Universities for Research in Astronomy, Inc., under cooperative agreement with the National Science Foundation.}   package. 
Second, a custom-designed pipeline developed with python and shell scripts was used to search for any new optical transients in error circle of radius 1.3 arcmin around the best position of FRB 20181130B by comparing with  
 astronomical catalogs including USNO B1.0 (Monet et al., 2013) with a typical limit magnitude of V=21 mag, Gaia dr2 (Gaia Collaboration et al., 2018) with a limit magnitude of G=18 mag, and Pan-STARRS DR1 (Chambers et al., 2016) whose typical limit magnitude is r$<$21.5 mag for single image or r$<$23.2 mag for stacked images.
All the above catalogues are deeper than the detection limit of each GWAC single image.
As shown in Figure.\ref{findchart}, none of any new credible optical transients were detected in  image \#0731 as well as in all other single fames.
  All  images also have been investigated by human eyes confirming the above conclusion of non-detection.  
  All these results are  shown in Table.\ref{limitmag} and Figure.\ref{limitmag_lc}.
In the upper panel in Figure.\ref{limitmag_lc}, the detection magnitude for each frame during our observations were displayed. The x-axis is the time in seconds relative to the event time, the y-axis is the 3 sigma limit magnitude. There is a global trend of the detection limit during the whole observations. The detection limit became deeper since the start observation due to the change of the background noise. Since the time about 3000 sec before the event time, the limit magnitude became relative stable.
  The below panel of Figure.\ref{limitmag_lc} shows the histogram of each limit magnitude shown in the upper panel.  As displayed with red dot line, a gaussian distribution fit the data well with  $\mu=15.43$ mag and $\sigma = 0.04$ mag. The Kolmogorov-Smirnov (K-S) test  was adopted for the fitting above yielding a result of  $p-value$ with 0.3303, which was larger than the critical value of 0.05 as a null hypothesis for a normal distribution. 
 This limit magnitude corresponds to a value of $\sim15.8$ mag in AB natural system.
after the considering the correction of the Galactic extinction of 0.108 mag (Schlafly \& Finkbeiner 2011) in R-band along the line of sight.

\begin{figure}[htbp]
\begin{center}
 \includegraphics[width=0.5\textwidth]{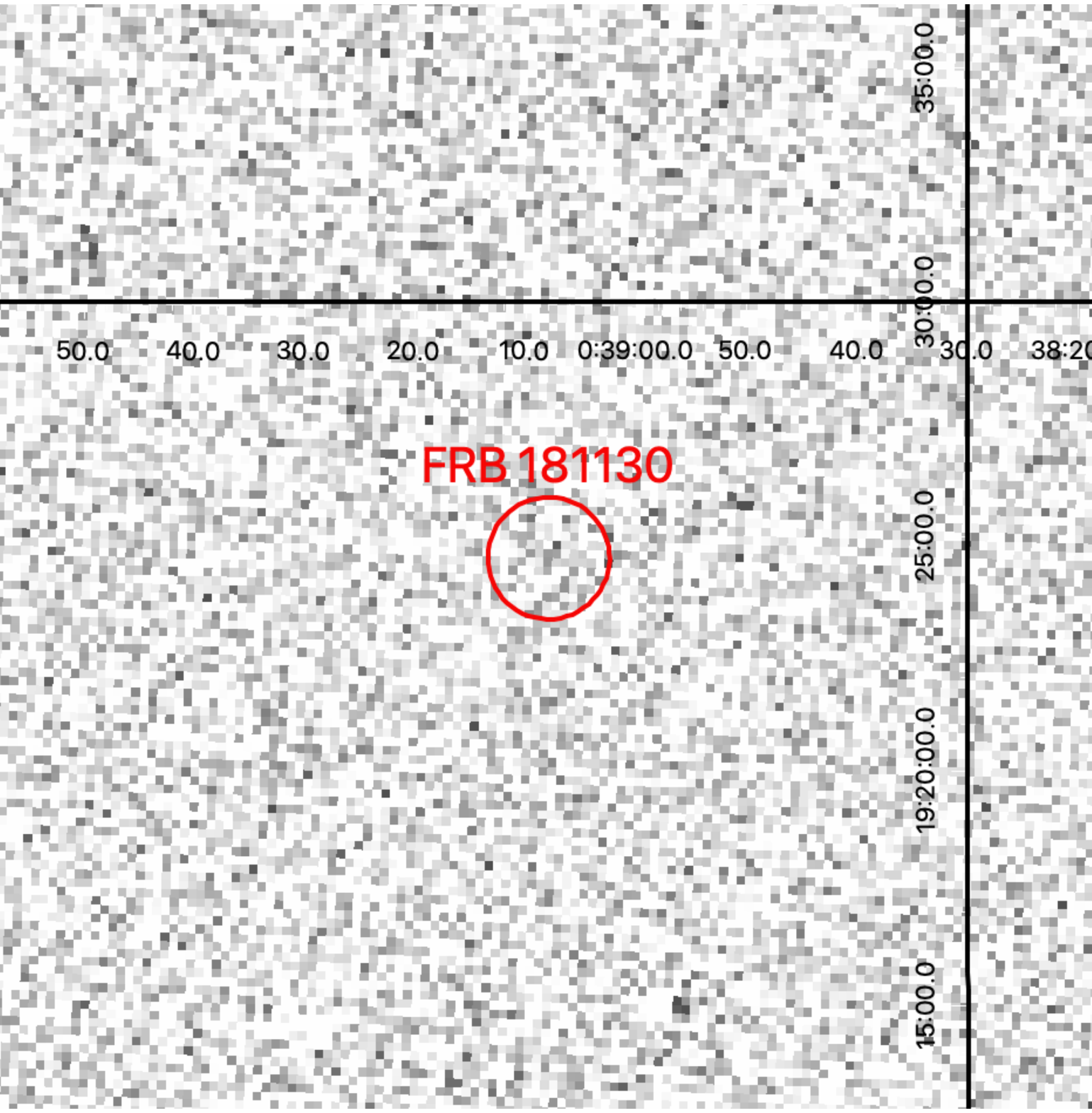}
\caption{A finding chart for subtraction between the consecutive images of \#0731 and \#0730 centering at the position of FRB 20181130B. The total size of this finding chart is about 11.5$\times$11.5 arcmin. The red circle shows the uncertainty of the position of FRB 20181130B with a radius of 1.5 arcmin. The top is the north. The right is the east. The grid of world coordinate system is displayed in black line. }
\label{subtraction}
\end{center}
\end{figure}

We also apply a difference image analysis via the hotpants package (Becker A., 2015)  to search for any new sources or variables during the burst time, taking the image \#0730 as a reference. 
Figure.\ref{subtraction} shows the residual image obtained by the subtraction between consecutive images \#0731 and \#0730. No any new credible source or variable was found in the residual image.

%\begin{figure}[htbp]

\iffalse
\begin{table}
\begin{center}
\caption{Properties of S1 (J003902.54+192431.5) derived from Gaia dr2 (Gaia Collaboration et al., 2018).}
\begin{tabular}{cccc} 
\hline 
RA & DEC &  G-band &  d \\
(J2000) & (J2000) &   mag & pc \\
\hline
00:39:02.523 &	+19:24:31.493 &   11.833 & 1.18542$\pm$0.0475 \\
\hline \label{Survey}
\end{tabular}
\end{center}
\end{table}
\fi

\begin{table}
\caption{Parts of GWAC observation log for FRB 20181130B. The filter is in white band. The exposure time was 10 seconds for each frame. $T_{0}$ is the burst time of FRB 20181130B derived from Niu et al., (2021). 
$T_{1}$ and $T_{2}$ are the expected times for optical emission associated to this event by considering the uncertainty of the dispersion measurement, respectively.
%$T-T_{1}$ is the expected delay time from optical emission to FRB derived from the wavelength of 1.5 GHz. 
%$T-T_{2}$ is the calculated delay time from optical emission to FRB derived from the wavelength of 1.16 GHz. 
All these magnitudes were in Vega system and  not corrected for the Galactic extinction of $A_R=0.108$ mag (Schlafly \& Finkbeiner 2011). }
\begin{center}
\begin{tabular}{cccccc}
\hline
 Start-time ($T$)   &   $T-T_{0}$  &   $T-T_{1}$  &  $T-T_{2}$ &  3$\sigma$   &  Image \\
   JD- 24584530.0   &             Second    &      Second    &   Second      &   Mag  & \\
  \hline
3.041501  &  -101.345  &  -98.2121  &  -98.1881  &  15.378  &  0725  \\
3.041675  &  -86.312   &  -83.1791  &  -83.1551  &  15.413  &  0726  \\
3.041849  &  -71.278   &  -68.1451  &  -68.1211  &  15.440  &  0727  \\
3.042022  &  -56.331   &  -53.1981  &  -53.1741  &  15.389  &  0728  \\
3.042196  &  -41.297   &  -38.1641  &  -38.1401  &  15.445  &  0729  \\
3.042370  &  -26.264   &  -23.1311  &  -23.1071  &  15.454  &  0730  \\
3.042543  &  -11.316   &  -8.1831   &  -8.1591   &  15.423  &  0731  \\
3.042717  &  3.717     &  6.8499    &  6.8739    &  15.485  &  0732  \\
3.042890  &  18.664    &  21.7969   &  21.8209   &  15.413  &  0733  \\
3.043064  &  33.698    &  36.8309   &  36.8549   &  15.430  &  0734  \\
3.043238  &  48.732    &  51.8649   &  51.8889   &  15.487  &  0735  \\
3.043411  &  63.679    &  66.8119   &  66.8359   &  15.465  &  0736  \\
3.043585  &  78.712    &  81.8449   &  81.8689   &  15.452  &  0737  \\
3.043758  &  93.660    &  96.7929   &  96.8169   &  15.441  &  0738  \\
\hline
\end{tabular}
\end{center}
\label{limitmag}
\end{table}

 \section{Results and Discussion}

There is a growing catalogue of the many different theories proposed for FRBs (Platts et al., 2019).  Any optical emission mechanisms and the brightness in optical band associated with FRBs are highly uncertain. 
In this work, we first focused on the search for prompt short-time (10ms-10 sec) optical emission associated with FRB 20181130B, and then also performed a search for long-duration transient.
All the optical flux limit and the corresponding AB magnitude derived in this work for each scenario  in the following discussion are summarized in Table.\ref{fluxlimit}.

%\subsection{Prompt optical pulse with $\sim 10$ milliseconds duration}
\subsection{Prompt optical emission}

We first consider scenarios where the optical emission has the same duration as the FRB. In these models (Yang et al. 2019), 
the single-zone inverse Compton (IC) scattering
in the pulsar magnetosphere model predicts the highest optical flux density and the optical counterpart having the same duration as the radio band. 
In this model, the radio radiation is produced by coherent curvature radiation of high-energy electrons, while the optical radiation is produced by IC scattering in the same region as the radio radiation.
With the equation (12) in Yang et al., 2019, 
$\frac{F_{\nu, \mathrm{IC}}}{F_{\nu, 0}} \simeq 5 \times 10^{-5} \eta_{\gamma} \times\left(\frac{\mu_{\pm}}{10^{3}}\right)\left(\frac{B}{10^{14} \mathrm{G}}\right)\left(\frac{P}{10 \mathrm{~ms}}\right)^{-1} $,
one could roughly estimate the
optical flux with radio flux based on the parameters of pulsars including magnetic strength $B$, the period of the pulsar $P$, the multipilicity $\mu_{\pm}$ resulting from  the electron-positron pair cascade (Yang et al., 2019), and the fraction of the electrons/positrons $\eta_{\gamma}$.  
For simplicity, assuming $\eta_{\gamma} = 1$,
using the extremely parameters with  $B\sim10^{15}$ Gauss, $P\sim1$ ms and  $\mu_{\pm} \sim 10^4$,  
the optical flux can be as high as  $F_{\nu,opt}\sim 5\times10^{-2} F_{\nu,radio}$. 
However, we also noted that if we use a typical values of a Galactic pulsar with $B=10^3$ Gauss, $P=1$ sec and $\mu_{\pm} \sim 10^3$, the optical flux would be about $F_{\nu,opt}\sim 5\times10^{-8} F_{\nu,radio}$.
In the case of FRB 20181130B, $F_{\nu, \mathrm{radio}}$ was reported as $\sim 20.6$ mJy (Niu et al., 2021). One could expect that the optical flux would be  $F_{\nu,opt}$ $\sim$1.0 mJy for extremely case or 
$1.0\times10^{-6}$ mJy for typical Galactic pulsar case.

Observationally for FRB 20121102A, the measured duration is 9.52$^{+5.94}_{-5.08}$ ms (Niu et al., 2021). 
The optical duration timescale shall be approximately around 10 ms.
 Following Yang et al., (2019), the optical flux density could be given with 
$F_{\nu, opt }=\left(\frac{T_{60}}{\tau_{\mathrm{ms}}}\right) 10^{\left(8.32-0.4 m\right)} \mathrm{Jy}$,
where  $\tau_{\mathrm{ms}}$ is the optical pulse duration in milliseconds and $T_{60}$ is the normalized exposure time of 60 seconds. 
With an optical pulse duration of $\sim 10$ ms, a GWAC exposure time of 10 s and the typical limit magnitude of 15.8 mag in AB natural system, 
the optical flux limit $F_{\nu, opt}$ would be deduced to be $\sim$1.66 Jy, which is  higher  than the optimistic estimate of the optical flux (1.0 mJy) by a factor of about 1660 or 8.0 magnitudes.

Other models (Yang et al., 2019) including IC scattering in one-zone emission from Masers in an outflow or in two-zone by Galactic energetic electrons, or the model for optical emission produced from the intrinsic mechanism of FRBs,  predict optical flux too low to be detected by GWAC-like facilities.
For example, 
under the consideration of one coherent mechanisms, curvature radiation by bunches
(Katz 2014, 2018; Kumar et al., 2017; Ghisellini \& Locatelli 2018; Yang \& Zhang 2018 ),
the optical flux is predicted by Yang et al., (2019) to be between 
$F_{\nu, \mathrm{opt}} \simeq\left(\nu_{\mathrm{opt}} / \nu_{\mathrm{radio}}\right)^{-1.6} F_{\nu, \mathrm{radio}}$ 
and 
$F_{\nu, \mathrm{opt}} \simeq\left(\nu_{\mathrm{opt}} / \nu_{\mathrm{radio}}\right)^{-(2 p+4) / 3} F_{\nu, \mathrm{radio}}$ with $p \geq 2$, where $p$ is the electron energy index.
Based on the above model, the predicted value of $F_{\nu, \mathrm{opt}}$ should be between $3.0\times10^{-11}$ Jy and $8.0\times10^{-21}$ Jy. This prediction of the brightness of the associated optical emission is fainter  of about 11 orders of magnitude and is a big challenge to be detected. 
 
On the other hand, as discussed by Hardy et al., (2017), the duration of any optical bursts can last up to five times wider than that of associated FRBs,  by assuming that the FRB mechanism follows similar behavior to the Crab pulsar, in which some optical pulses have been detected (Shearer et al., 2003; Slowikowska et al., 2009; Mignani 2010; Shearer et al., 2012).  If it is the case for FRB 20181130B,  the optical emission duration can be as long as 50 ms. The optical flux limit $F_{\nu,opt}$ would be estimated to be 0.33 Jy. 

\begin{figure}[htbp]
\begin{center}
 \includegraphics[width=0.5\textwidth]{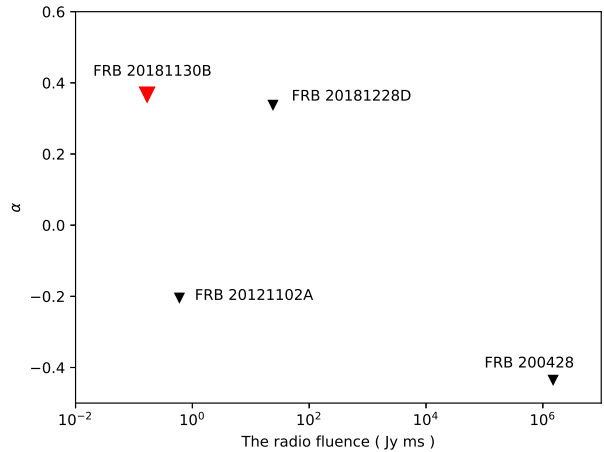}
\caption{The level of constraints for the spectral indices $\alpha$ from optical to radio wavelengths for four FRBs with their radio fluence. The four FRBs are FRB 200428 (Lin et al., 2020), FRB 20181228D (Tingay \& Yang 2019; Farah et al., 2019), FRB 20121102A (Hardy et al., 2017) and FRB 20181130B in red triangle studied in this work. The detailed data are summarized in Table.\ref{opt_radio_ratio}. }
\label{spectral_RadioFluence}
\end{center}
\end{figure}

By multiplying the flux density limit of 1.66 Jy by the duration of the optical emission of 10 ms, we  obtained the maximum simultaneous optical fluence of 16.6 Jy ms for FRB 20181130B.   
Assuming that the emission in the optical and radio frequencies had the same intrinsic source,  the 
broad-band spectral slope $\alpha$ ($f_\nu \propto \nu^{\alpha}$)  from optical to radio wavelengths for FRB 20181130B is smaller than 0.367. 
With the same method, we also calculate other three FRBs whose prompt optical observations were obtained in the literature. They are FRB 200428\footnote{There is no official name in the Transient Name Server to date.}, (Lin et al., 2020) FRB 20181228D (Tingay \& Yang 2019; Farah et al., 2019) and FRB 20121102A (Hardy et al., 2017). 
All the results are summarized in Table.\ref{opt_radio_ratio} and displayed in Figure.\ref{spectral_RadioFluence}.
The constraints on the spectral slope  $\alpha$ for FRB 20181130B  in this work is comparable to that of  FRB 20181228D (Tingay \& Yang 2019; Farah et al., 2019), but  shallower than the limit of FRB 200428 (Lin et al., 2020) or FRB 20121102A (Hardy et al., 2017). 
 
\subsection{Long duration optical emission}

The two-zone IC scattering process may produce 
optical emission with a duration longer  than that of the associated FRB 
because the scattering region could be much larger than the  FRB emission region
(Yang et al., 2019;  Tingay \& Yang et al., 2019). 
One of  the long-duration optical emission model is the pulsar nebula (e.g., SNR) model (Yang et al., 2019), in which  the optical flux can reach up to $F_{\nu,\mathrm{opt}}\sim 8.8\times10^{-3} F_{\nu, \mathrm{radio}}$.  With their prediction, $F_{\nu,\mathrm{opt}}$ can 
be derived as high as 0.18 mJy with  $F_{\nu, \mathrm{radio}}$ of  $\sim 20.6$ mJy (Niu et al., 2021).

Assuming that the duration of the optical emission is equal to or much longer than the GWAC single exposure time, 
the optical flux density can be constrained to 1.74 mJy 
for a GWAC single exposure(10 sec). 
which is still about one order of magnitude shallower than the predicted value of the model above. 
On the other hand, 
since the predicted duration may be as long as $10^{4}$ seconds according to the pulsar nebula (e.g., SNR) model (Yang et al., 2019),
 we investigate the possible  longer duration optical emission by stacking the 405 images after the eruption times. The observation duration for these images were from 13:01:14 UT to 14:42:15 UT on the same night.  The total coverage time is 6060 sec and the effective exposure time is 4040 seconds. There is no any new source around the best position (Niu et al., 2020) down to  3$\sigma$ upper limit magnitude of 17.8 mag in R band or 18.39 mag in AB system, calibrated to the SDSS catalogues\footnote{http://www.sdss.org/dr6/algorithms/sdssUBVRITransform.html\\\#Lupton2005 (R = r - 0.2936*(r - i) - 0.1439;  sigma = 0.0072) }.
This upper limit magnitude corresponds to 0.16 mJy, which is close to the predicted maximum  discussed above.

\begin{table}
\begin{center}
\caption{Summary of the constraints for optical brightness for different durations. The AB magnitude in fourth column is derived with the equation $m=-2.5 \log _{10}\left(F_{\nu} / 3631 \mathrm{Jy}\right)$ where F$_{\nu}$ is the value in the third column for each line. }
\begin{tabular}{ccccc} 
\hline 
Mid-time & Optical duration & Flux limit & AB magnitude & Images\\
(second) &                           &                &               & \\
\hline
0 & 10 ms & 1.66  Jy  & 8.35 & single\\
0 & 50 ms & 0.33 Jy  & 10.10 & single \\
0  & $\geq$10 s  &  1.74 mJy & 15.80  & single \\
3030.5  & 6061 s & 0.16 mJy & 18.39 &stacked \\
\hline \label{fluxlimit}
\end{tabular}
\end{center}
\end{table}

\begin{table}
\begin{center}
\caption{Broad-band spectra slope ($f_\nu \propto \nu^{\alpha}$)  from optical to radio wavelengths during the prompt phase. Note that optical frequency for all events adopted here is set to the same which has a negligible impact on the our constraints. }
\begin{tabular}{cccccccc} 
\hline 
 ID & Opt. Fluence  & Radio Fluence  & $\nu_{opt}$ & $\nu_{radio}$ &  $\alpha$ & Event & Reference \\
      &           (Jy ms) &  (Jy ms) &  (Hz)  &   (Hz)  &    &   &   \\   
\hline 
  1   & 16.6         &         0.168      &    4.0e14    &       1.5e9  &  0.367  &  FRB 20181130B              & This work \\                                                                                                                     
  2   &  4.4e3       &          1.5e6     &     4.0e14   &        0.6e9  &  -0.435  &  FRB 200428   & Lin et al., 2020   \\
  3  &  $>$2000          &         $>$24      &       4.0e14    &       0.835e9 & 0.338 &   FRB 20181228D & Tingay \& Yang 2019; \\
     &                     &                     &                  &                  &         &                          & Farah et al., 2019\\
  4   & 0.046        &         0.6      &      4.0e14      &     1.36e9 & -0.204 &  FRB 20121102A  & Hardy et al., 2017 \\
\hline \label{opt_radio_ratio}
\end{tabular}
\end{center}
\end{table}

\subsection{Optical emission associated with future local FRBs}

Many efforts have been devoted to search for multi-wavelength counterparts (see the recent review of Nicastro et al., 2021). There are generally three strategies: 1) standard triggered follow-up observations(e.g., Petroff et al., 2015; Bhandari et al. 2018), similar to the observation strategy for optical afterglow of gamma-ray bursts; 2) target search for repeating FRBs (e.g., Scholz et al., 2016, 2017; Hardy et al, 2017) or periodic FRBs (e.g., Kilpatrick et al., 2021); 3) simultaneous observations with wide-field telescopes (e.g., Tingay \& Yang 2019).

Here we presented a  simultaneous observation by GWAC covering the entire period of the expected optical emission of FRB 20181130B after taking into account its high dispersion measurements. Our analysis shows that no any optical emission was detected in the single image or stacked image. 
The estimated distance  of FRB 20181130B is z$\sim$1.83 (Niu et al., 2021), which makes the event to be by far one of the highest redshift in the FRB catalogue.  
The radio peak fluxes and fluence make FRB 20181130B located in the  faint part of the distribution of distance and observed fluence (see Figure 2 of The CHIME/FRB Collaboration 2020).

However,
one shall note that the average value for most known FRBs is at the level  of about 10 Jy ms (e.g., Shannon et al., 2018; The CHIME/FRB Collaboration 2020).
Most of the detected FRBs are at a distance of about $10^{8}-10^{10}$ pc. 
Theoretically,  the short duration optical flashes accompanying these typical FRBs are still be expected to be detected with future efforts. 
 For example, 
 bright optical flash with a time scale of 1 sec is expected  proposed by Beloborodov (2020) with the model that the blast wave
 strikes the wind bubble in the tail of a preceding flare in frequent repeaters, though the expected rate is a small fraction
 of the FRB rate. Considering the uncertainty of the optical emission predicted by various of models (eg., Platts et al, 2019; Nicastro et al., 2021), observations in the future to search and validate the optical emission associated to the FRBs would be optimized in the following aspects:  1) faster optical cadence down to subsecond timescale; 2) deeper detection capability for one exposure time with large-aperture telescopes; 3)wide-field field of view; 4) commensal observing between optical and radio facilities; 5) multi-telescope monitor the same sky located at a long distance to distinguish the contamination from cosmic ray, artificial objects or other instrument defects; 6) target monitoring for those repeaters or periodic events.
Among those FRBs, 
the most anticipated detection is for those nearby (or low-DM) and brighter FRBs, such as the energetic FRB 20180110A with a fluence of about 390 Jy ms, 
or the brightest radio burst from SGR 1935+2154 
($\sim$220 kJy ms, The CHIME/FRB Collaboration 2020) with a distance of 9.5 kpc  (Bochenek et al. 2020), 
or the recently reported repeating  FRB 20200120E (Bhardwaj et al., 2021) which was found to be from M81 at 3.6 Mpc.

Given the high sensitivity of FAST, CRAFTS  tends to detect more distant and fainter FRBs (e.g., Zhang et al., 2018; Niu et al., 2020). However, it is still anticipated that some of  bright, nearby FRBs may be detected. 
GWAC is located at Xinglong Observatory in China, which is near the location of the FAST. Due to the similar visibility of the sky, the observation field of GWAC can always be selected to cover the same field of view of FAST in order to observe the same sky simultaneously.
Furthermore, GWAC has a plan to upgrade parts of  cameras equipped with complementary metal oxide semicon (CMOS)  by increasing the temporal resolution to 1 sec, which is more advantageous for detecting short timescale transients ($<$1 sec) such as fast optical emission associated with FRBs, by decreasing the background noise and shortening the dead time greatly compared to the CCD used currently.  If some bright nearby FRBs were detected by FAST, then the high temporal resolution and simultaneous observations of GWAC will detect the associated fast optical bursts or make better constraints on their brightness. 

On the other hand, 
if counterpart events were rare in the local universe, with characteristics of  high cadence, large FoV and long-term operation,  a blind search by GWAC can also increase the detectability of  short-living optical counterparts, although its detection capability is shallow. Similar discussion for the future searching strategy is also presented by Chen et al., (2020). For example,
taking the all-sky event rate of around 250 per day estimated for FRBs like the Lorimer-burst (Lorimer et al., 2007), 
about 1.3 bursts would be seen in the GWAC field each hour after the full system have been set with the FoV of 
about $\sim$5000 $deg^{2}$  (Wei et al., 2016). 
Following the  discussion by Lyutikov \& Lorimer (2016), the optical brightness would be expected to be 12.6 mag or 9.72 mag for a 15 sec or 1 sec cadence, respectively,  by adopting the parameter of the  radio peak flux density $F_{Jy}$=30, the optical duration $\tau=15$  ms and the ratio between optical flux and radio flux $\eta = 1$. Such brightness shall be detectable by GWAC system.

\section{Summary}
In this paper, we report on synchronized observations with GWAC during the outburst time of FRB 20181130B. 
There is no new source in the error circle around the best position in either single or stacked images. 
The optical flux limit obtained by our observations are 1.66 Jy (8.35 mag), 0.33 Jy (10.10 mag), 1.74 mJy (15.80 mag) and 0.16 mJy (18.39 mag) for optical emission duration of 10ms, 50ms, 10 s and 6060 s, respectively. 
All images also have been investigated by eye confirming the above conclusion of non-detection.
Due to the nature of FRB 20181130B with a long distance and low radio fluence, the optical fluxes predicted by the FRB model for 
the fast optical bursts (10 ms to 10 s) are too low to be well constrained by our observations. However, the optical flux limit after stacking GWAC images is almost comparable to the maximum level of long-duration optical emission predicted by pulsating nebulae (e.g., SNR) models (Yang et al. 2019). We also discuss the characteristics of the most  FRBs detected to date. If the GWAC system is updated in the future and the observing strategy is optimized by overlaying the same sky field  with that of the FAST, it is highly anticipated that better constraints on the optical brightness associated to nearby bright FRBs could be derived.

\section{Acknowledgement}
The authors thank the anonymous referee for a careful review and helpful suggestions that improved the manuscript. 
The authors thank Bing Zhang and Yuanpei Yang for discussions for this work. 
This study is supported from the National K\&D
Program of China (grant No. 2020YFE0202100)
 and the  National Natural Science Foundation of China (Grant No.
11973055, U1831207, 11863007). This work is supported by the Strategic Pioneer Program on Space Science, 
Chinese Academy of Sciences, grant Nos. XDA15052600 \& XDA15016500, 
 and  by the Strategic Priority Research Program of the Chinese Academy of Sciences, Grant No.XDB23040000. 
 XLP is partially supported by the Research Program Nos. CMS-CSST-2021-B1. 
YGY is supported by the National Natural Science Foundation of China under grants 11873003. JW is supported by the National
Natural Science Foundation of China under grants 11473036 and 11273027.
DZG is supported by the National Key Research and Development Program of China (grant No. 2017YFA0402600), the National SKA Program of China (grant No. 2020SKA0120300) and the National Natural Science Foundation of China (grant No. 11833003).
This work has made use of data from the European Space Agency (ESA) mission
{\it Gaia} (\url{https://www.cosmos.esa.int/gaia}), processed by the {\it Gaia}
Data Processing and Analysis Consortium (DPAC,
\url{https://www.cosmos.esa.int/web/gaia/dpac/consortium}). Funding for the DPAC
has been provided by national institutions, in particular the institutions
participating in the {\it Gaia} Multilateral Agreement.
This research has made use of the VizieR catalogue access tool, CDS, Strasbourg, France (DOI: 10.26093/cds/vizier). The original description of the VizieR service was published in A\&AS 143, 23

\end{document}